\documentclass[12pt]{iopart}
\usepackage[english]{babel}
\usepackage[utf8x]{inputenc}
\usepackage[T1]{fontenc}
\usepackage{braket}
\usepackage{graphicx}
\usepackage{multirow}
\usepackage{booktabs}

\usepackage{appendix}

\usepackage{indentfirst}
\usepackage[export]{adjustbox}
\usepackage{subcaption}

\usepackage{lipsum}
\usepackage{cite}
\expandafter\let\csname equation*\endcsname\relax
\expandafter\let\csname endequation*\endcsname\relax
\usepackage{amsmath}
\usepackage{amssymb}
\usepackage{marginnote}
\usepackage{graphicx}
\usepackage{dirtytalk}

\usepackage{pifont}
\usepackage{amssymb}
\usepackage{bm}
\usepackage{amsfonts}
\usepackage{amsmath}
\usepackage{amssymb}

\def\prl{Phys. Rev. Lett.}

\def\prd{Phys. Rev. D}

\begin{document}

\title[\,]{De Sitter-like configurations with asymptotic quintessence environment}

\author{Roberto Giamb\`o$^{1,2,3}$ and Orlando Luongo$^{1,2,4,5,6}$}
\address{$^1$Universit\`a di Camerino, Via Madonna delle Carceri 9, Camerino, Italy.}
\address{$^2$Istituto Nazionale di Fisica Nucleare, Sezione di Perugia, via A. Pascoli, I-06123 Perugia, Italy.}
\address{$^3$IAPS, INAF - Tor Vergata, 00133 Roma, Italy.}
\address{$^4$SUNY Polytechnic Institute, 13502 Utica, New York, USA.}
\address{$^5$INAF - Osservatorio Astronomico di Brera, Milano, Italy.}
\address{$^6$Al-Farabi Kazakh National University, Al-Farabi av. 71, 050040 Almaty, Kazakhstan.}

\ead{orlando.luongo@unicam.it, roberto.giambo@unicam.it}

\begin{abstract}
We examine a spherically-symmetric class of spacetimes carrying vacuum energy, while considering the influence of an external dark energy environment represented by a non-dynamical quintessence field. Our investigation focuses on a specific set of solutions affected by this field, leading to  distinct kinds of spacetime deformations, resulting in regular, singular, and wormhole solutions. We thoroughly discuss the underlying physics associated with each case and demonstrate that more complex deformations are prone to instability. Ultimately, we find that our results lead to an \emph{isotropic de Sitter-like solution} that behaves as a quintessence fluid. To achieve this, we investigate the nature of the corresponding fluid, showing that it cannot provide the sound speed equal to a constant equation of state {parameter} near the center. Consequently, we reinterpret the fluid as a slow-roll quintessence by investigating its behavior in asymptotic regimes. Further, we explore the potential implications of violating the isotropy condition on the pressures and we finally compare our findings with the de Sitter and Hayward solutions, highlighting both the advantages and disadvantages of our  scenarios.
\end{abstract}

\section{Introduction}\label{introduzione}

The evidence in support of cosmic acceleration is a an established fact of modern cosmology  \cite{Copeland:2006wr}. The cosmological constant, $\Lambda$, is commonly considered the simplest approach to describe the large-scale dynamics of the universe. Specifically, it appears as a key ingredient in the standard \emph{$\Lambda$CDM paradigm}\footnote{Dark energy models lie on assuming the cosmological principle to hold. Alternatives that aim at violating it are also possible, see e.g. \cite{Luongo:2021nqh,Shamir:2022wlx, Krishnan:2022qbv}.}.

Within the cosmological puzzle, $\Lambda$ is interpreted in terms of a constant vacuum energy term  \cite{Perivolaropoulos:2021jda} derived from quantum fluctuations \cite{Martin:2012bt}. According to this ansatz, the cosmological constant represents a form of dark energy that remains unchanged over time\footnote{For the sake of completeness, it is remarkable to emphasize a significant distinction between the microphysics of dark energy and $\Lambda$. While the $\Lambda$CDM model is statistically favored, it remains theoretically incomplete, mainly due to the cosmological constant problem \cite{weinberg1989cosmological}. Addressing this issue would offer new insights into the validity of the $\Lambda$CDM scenario  \cite{Luongo:2018lgy, Luongo:2014nld, Belfiglio:2023rxb, Belfiglio:2023eqi}.}. Accordingly, the $\Lambda$CDM framework aligns well with observational data, albeit ongoing debates continue regarding recent tensions and conceptual challenges within the model itself, see e.g. \cite{Verde:2013wza}. Nevertheless, dark energy is commonly described through barotropic fluids, exhibiting the onset of cosmic speed up at a given transition time \cite{Capozziello:2021xjw,Farooq:2013hq,Capozziello:2014zda,Farooq:2016zwm,Muccino:2022rnd,Alfano:2023evg}.

A promising alternative to barotropic fluids, commonly employed to describe dark energy scenarios, involves the use of scalar fields. They represent effective fluids that differ significantly from barotropic approaches due to their distinct underlying physics, see e.g.  \cite{Linder:2008ya,Giambo2024,Capozziello:2019cav}. The simplest approach,  employing a positive kinetic term along with a unspecified potential, is often referred to as \emph{quintessence}. Quintessence can reproduce the effects of dark energy mimicking the cosmological constant in the slow-roll regime\footnote{For a criticism to  the standard quintessence paradigm, see e.g. \cite{2022CQGra..39s5014D,2010PhRvD..81d3520G,Chimento:2010vu,Chimento:2007da}.}. Indeed, assuming $P$ and $\rho$ the cosmological total pressure and density respectively, $w\equiv\frac{P}{\rho}=-1$  exactly represents the equation of state {parameter} for the cosmological constant case, having instead  $w\simeq-1$ for the quintessence slow-roll approximation, in which the kinetic energy is subdominant than the potential.

In addition, very recently the study of black holes has gained significant importance due to the modern discoveries of gravitational waves and black hole shadows \cite{Abbott2016,EventHorizonTelescope:2019dse}. Consequently, investigating the behavior of black holes within quintessence fields or, more generally within dark energy environments, can provide valuable insights into the interaction between black holes and dark energy \cite{Zhang:2023neo}. Initial attempts incorporate black holes into quintessence environments \cite{Kiselev:2002dx}, albeit yielding dark energy fluids that exhibit  distinct radial and tangential pressures \cite{Visser:2019brz}, depending on the radial coordinate, $r$, \textit{i.e.},  disagreeing with the  cosmological principle that assumes identical radial and tangential pressures, functions of cosmic time only.

Moreover, recent advancements have revealed the possibility of non-singular black hole configurations, predicting the theoretical existence of regular black holes \cite{Lan:2023cvz,Torres:2022twv,Malafarina:2022wmx}. These solutions, derived from Einstein's field equations, introduce notable enhancements to the properties of black holes \cite{Luongo:2023aib,Boshkayev:2023fft,Luongo:2023jyz,Boshkayev:2023rhr}. One intriguing characteristic of regular black holes is their ability to carry  vacuum energy. For example, this scenario has been extensively explored through the study of the Hayward solution \cite{Hayward:2005gi} and its extensions \cite{Mosani:2023awd,Bambi2013PhLB,Zhou:2023lwc,Zhang:2014bea}. There, the fundamental requirement is the presence of a de Sitter phase at small radii, transitioning to a Schwarzschild solution at very large distances, far from the conventional singularity found at the origin of radial coordinates.

Motivated by these considerations, our work delves into a class of regular solutions by imposing a spherically-symmetric spacetime with a  $00$-component, exhibiting a de Sitter-like behavior. To accomplish this, we solve  Einstein's equations while demanding isotropic pressures in fulfillment to dark energy scenarios. Specifically, we first consider that vacuum energy can be function of an external non-dynamical field that \emph{deforms} the de Sitter solution modifying the mathematical structure of the metric itself. Thus, we reinterpret the field as a possible external temperature induced by the dark energy environment, in agreement with recent developments toward thermodynamic black holes \cite{Capozziello:2022ygp}. To guarantee that the metric is Lorentzian and remains regular at all radii, we find that our solution accounts for a (quasi) anti-de Sitter cosmological constant contribution characterized by negative energy density and positive pressure, whose  resulting equation of state  asymptotically resembles a quintessence behavior. Further, we investigate the main thermodynamic properties of our solution, involving the study of the sound speed at both small and large radii. To this end, we single out opportune options for our free parameters, appropriately adjusted to reproduce regular and singular solutions. Within this treatment, we explore the possibility of mimicking a wormhole, comparing our solution with early literature. Last but not least, we investigate the case in which pressures are not isotropic, studying either the case of vanishing radial pressure or zero tangential one. Again, the properties of transporting vacuum energy are investigated, showing the main differences with respect to the isotropic case and, particularly, showing instabilities into the computed sound speeds. Regarding deformation, we emphasize that it does not delve into a class of interesting candidates to transport vacuum energy. Physical consequences of this aspect are thus described in detail. Finally, our solution is compared with the Hayward and de Sitter spacetimes, emphasizing the main departures and the regimes in which our spacetime appears better-behaving than those metrics.

The paper is structured as follows. in Sect. \ref{sezione2}, the spherical symmetric setup is developed and the way we construct our solution is introduced. In Sect. \ref{sezione3}, we describe a varying effective cosmological constant with no deformation, while in Sect. \ref{sezione4} the deformation is accounted. The physical interpretation of our fluid is reported in Sect. \ref{sezione5} and, finally, we discuss conclusions and perspectives in Sect. \ref{sezione6}.

\section{Setup in spherical coordinates}\label{sezione2}

The generic static spherical spacetime is
\begin{equation}
ds^2 =  -f(r) dt^2 + \frac{1}{g(r)} dr^2 + h(r) d \Omega^{2}\,,
\label{sferico0}
\end{equation}
where  $f(r), g(r)$ and $h(r)$ depend on the radial coordinate only. The stationary region, i.e.  where $\partial_t$ is a \textit{timelike} Killing field, corresponds to the region where those three functions are all positive.

In the coordinate system $(t,r,\theta,\varphi)$, the stress-energy
tensor for a perfect fluid is
\begin{equation}
\left[ T_{\nu}^{\mu} \right] =
\left[\begin{array}{cccc}
-\rho &       &      &     \\
      & P_{r} &      &     \\
      &       & P_{t} &     \\
      &       &      & P_{t}
\end{array}
\right] \,\,,
\label{emt}
\end{equation}
where we distinguish the radial pressure, $P_r$, from the tangential pressure, $P_t$.

Our purpose is to incorporate into the above metric  an \emph{external field, $\phi$},  that  does not exhibit dynamics though. Clearly, the simplest option involves the use of a scalar field describing a temperature related to the environment resulting from the presence of dark energy. Specifically, dark energy naturally acts as a thermal bath, effectively modifying the energy-momentum tensor. Notably, significant examples of such scenarios have been studied within the framework of external gases of dark energy surrounding black holes, see e.g.  \cite{Rajagopal:2014ewa,Capozziello:2022ygp}.

Thus, in order to deform  Eq. \eqref{sferico0}, we introduce the following modification
\begin{equation}
ds^2 =  -f(r,\phi) dt^2 + \frac{1}{g(r,\phi)} dr^2 + h(r) d \Omega^{2}\,,
\label{sferico2}
\end{equation}
where conventionally we hereafter take  $h(r)=r^2$, \textit{i.e.}, leaving unaltered under the action of deformation the three-volume associated with the spacetime itself\footnote{In general, any deformation on $h(r)$ can be removed through a change of coordinates, in order to reobtain the usual three-volume of a sphere. }. Here, the Schwarzschild coordinates are not fully-established since \emph{a priori} $f(r,\phi)\neq g(r,\phi)$.

The physical meaning of $f(r,\phi)$ implies that we search for an external field, $\phi$, that can deform the mathematical structure of our solution. In other words, $\phi$ is interpreted as an external field that modifies the energy-momentum tensor of our metric and, then, can act as an \emph{environment fluid}, whose properties resemble those of a temperature. Consequently, assuming that $\phi$ acts as a temperature, we can vary it expecting that as the system reaches the equilibrium, the physical properties of our metric would change accordingly. Hence, the presence of $\phi$ defines a class of metrics embedded in a temperature field, whose properties can significantly change the physics associated with the free parameters of our metric, as we will clarify throughout the text.

We proceed to include the external field motivated by the following physical arguments.
\begin{itemize}
    \item[-] The action of the field  is to switch the free parameters of our fluid, implying a phase transition that induces a modification of the spacetime itself.
    \item[-] By considering dark energy as the source for the external field, we find that the most general solution for the equation of state will also be dependent on both $r$ and $\phi$. As a result, the system is not in equilibrium. To restore thermal equilibrium, we can explore the asymptotic regime, where the dependence of $w$ on the radial coordinate disappears, and equilibrium can be restored.
\end{itemize}

On the other hand, Eq. \eqref{sferico0} with $h(r)=r^2$ can be recast as
\begin{equation}
ds^2 =  -\exp^{-2\Phi(r)} dt^2 + \left(1-\frac{2M(r)}{r}\right)^{-1} dr^2 + r^2 d \Omega^{2}\,,
\label{sfericoagain}
\end{equation}
where the functions, $\Phi(r)$ and $M(r)$, are related to the gravitational Newtonian potential and to the total mass generating the solution, respectively.

Consequently, deforming Eq. \eqref{sfericoagain} with $\phi$ implies that the potential is changed by virtue of thermal energy associated with the field. In other words, a possible deformation, that gives rise to Eq. \eqref{sferico2}, occurs when the potential and mass change according to

\begin{subequations}
    \begin{align}
\Phi(r,\phi)&=-\frac{1}{2}\log[f(r,\phi)],\label{phiii}\\
M(r,\phi)&=\frac{r}{2}\left(1-g(r,\phi)\right)\,.\label{MM}
    \end{align}
\end{subequations}

\begin{figure*}[h]
  \centering
    \includegraphics[width=.5\linewidth]{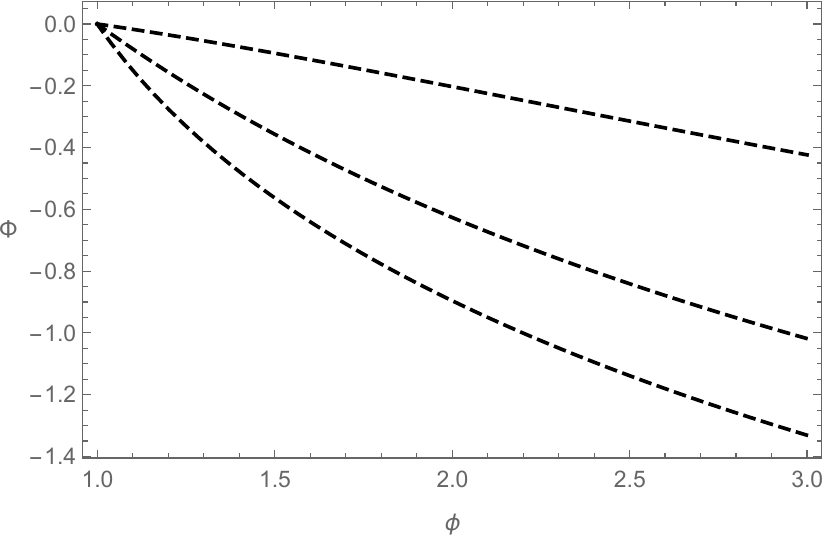}
  \hspace{0mm}\\
    \includegraphics[width=.5\linewidth]{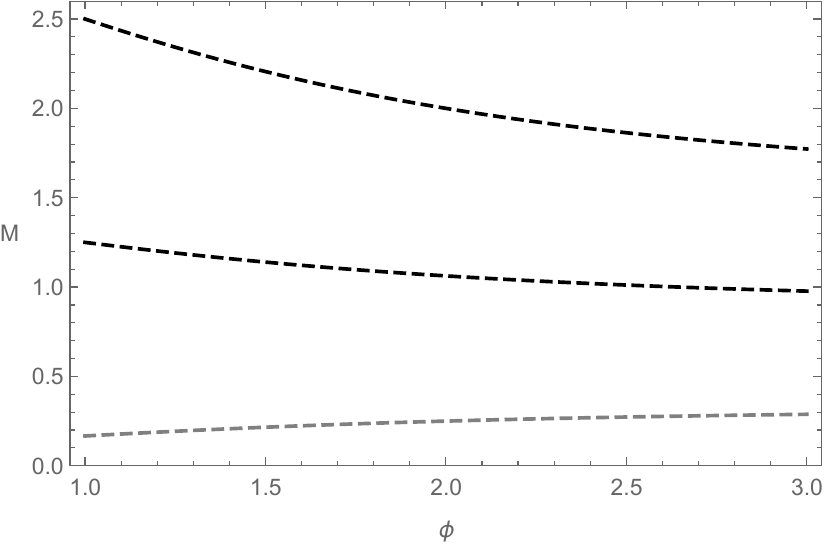}
    \vspace{7mm}\\
  \caption{Modifications induced by our solution with $\epsilon=0$ on $\Phi$ (top figure) and $M$ (bottom figure) got from Eqs. \eqref{phiii}-\eqref{MM} with indicative  $\Lambda_B={\frac12}$. For $\Phi$, we assume $r=1$ and we vary the field as $\phi\in[1;3]$, letting $\Lambda$ be  negative, having fixed $\phi_c=1$. The black dashed lines correspond to $\Lambda_B,5\Lambda_B$ and $10\Lambda_B$. For $M$, we fixed $\Lambda_B$ and vary $k_0=\{-10\Lambda_B,-5\Lambda_B\}$. In this case, the gray line corresponds to $k_0=-\frac{2}{3}\Lambda_B$.}
  \label{PhiandM}
\end{figure*}

The functional beahaviors of $\Phi$ and $M$ will be studied later in the text (see Fig. \ref{PhiandM}).

In particular, we notice that a generic change of coordinates may  transform the metric into an equivalent one,  removing $M(r,\phi)$. This appears possible since $M$ is a radial function and $\phi$ remains unaltered after a change of coordinates. Further, an environment field of dark energy can modify the gravitational field around the spherical without modifying the total mass that generates the solution itself.

Bearing this in mind, we focus below on the simplest approach that carries on vacuum energy, without assuming $f=g$ and regularly behaving at $r=0$. We then assume that vacuum energy sign can change under the action of $\phi$ and investigate which physical consequences can be argued from this recipe.

\subsection{Introducing vacuum energy}

Writing the Einstein equations as
\begin{equation}
G_\nu^\mu\equiv R_{\nu}^{\mu} - \frac{1}{2} R \, \delta_{\nu}^{\mu}
= 8\pi T_{\nu}^{\mu} \,\,,
\label{einstein_equations}
\end{equation}
with $G_\nu^\mu$ the Einstein tensor, $T^\mu_\nu$ the energy momentum tensor and $\delta^\mu_\nu$ the Kronecker delta, the simplest regular solution that transports vacuum energy is the de Sitter spacetime, where
\begin{equation}
f=g=1-\frac{\Lambda}{3} r^2\,,
\end{equation}
whose energy-momentum tensor is
\begin{equation}
T^{\mu}_\nu=-\Lambda\delta^\mu_\nu\,.
\end{equation}

To incorporate the presence of a deforming external field, $\phi$, the first conditions that can be considered are
\begin{subequations}
    \begin{align}
\Lambda&=\Lambda(\phi)\,,\\
g(r,\phi)&\neq f(r,\phi).\,
    \end{align}
\end{subequations}

In addition, we propose a second deformation, that includes the first as a limiting case, having
\begin{equation}\label{secondadeformazione}
f(r,\phi)= 1-\frac{\Lambda(\phi)}{3}r^{2+\epsilon(\phi)}\,,
\end{equation}

The physical interpretation of Eq. \eqref{secondadeformazione} comes from the de Sitter spacetime. Indeed, we intend to deform a standard configuration that carries on a constant form of energy, interpretable in terms of vacuum energy, namely the standard de Sitter configuration. To deform it, the simplest assumption is to depart from the $\propto r^2$ functional behavior of the solution, assuming that $\Lambda$ is no longer a constant. To depart from the $\propto r^2$ functional form, we thus introduce a deforming term, $\epsilon(\phi)$, that might depend on the external field, $\phi$, responsible for the transition, that acts on the spacetime itself.

Hence, assuming $\epsilon\ll1$, the corresponding approximation reads,
\begin{eqnarray}
f(r,\phi)\simeq -\frac{\Lambda(\phi)}{3}r^2(1+\epsilon\ln r)\,,
\end{eqnarray}
that, in the case of slightly evolving $\Lambda(\phi)\simeq \Lambda_B$, gives a logarithmic correction to the de Sitter spacetime that can be relevant to precise intervals of distances,  depending on the choice of $\epsilon(\phi)$. For example, assuming that $\epsilon\rightarrow0$ as $r\rightarrow0$ or $r\rightarrow\infty$ permits to neglect departures from a genuine de Sitter cosmological constant at asymptotic regimes, giving rise to corrections that arise only within the more internal intervals of radii.

Consequently, $\epsilon$ is physically the simplest correction possible to the de Sitter power-law dependence.

In view of the above interpretation, Eq. \eqref{secondadeformazione} can therefore show significant departures from the de Sitter case, that will be clarified later, as we introduce the asymptotic quintessence behavior.

In both the deformations, the equation of state {parameter} depends on $r$ and $\phi$.

For the sake of completeness, the influence of the external field, $\phi$, can also be encapsulated in postulating the sign of $\Lambda$ by hands. Indeed, as we will elucidate later, the sign of $\Lambda$ can be arbitrarily set to either positive or negative, resulting in vastly different properties for our solution. However, the rationale behind choosing one sign for $\Lambda$ over the other is grounded in existing literature where external fields can act as \emph{effective fluids}, altering the dynamics of the spacetime free terms, see e.g. \cite{Shaymatov:2020bso,Capozziello:2022ygp,Jusufi:2022jxu,Jawad:2023ypn,Oubagha:2023ghx}. Additionally, $\Lambda$ is associated with vacuum energy, and it is a well-known  that the sign of vacuum energy gives rise to de Sitter or anti-de Sitter phases, each exhibiting distinct physical properties, with recognized implications in quantum field theory \cite{Hubeny:2014bla}.

The negative cosmological constant plays a pivotal role in the anti-de Sitter/conformal field theory  correspondence, i.e., establishing a duality between a quantum gravity theory in anti-de Sitter space and a conformal field theory  in one dimension lower. This correspondence provides a unique avenue for exploring strongly coupled field theories by utilizing classical gravity\footnote{This clearly offers a valuable approach to investigating non-perturbative phenomena that elude traditional perturbative methods.}.

To this regard, one notable implication of the negative cosmological constant is, for example, the emergence of a holographic screen at the anti-de Sitter boundary. This screen acts as a boundary for anti-de Sitter space and encodes essential information about the bulk geometry. Essentially, the holographic screen serves as a holographic representation of the three-dimensional bulk, capturing its dynamics and allowing researchers to glean insights into the behavior of the underlying quantum field theory.

In essence, the correspondence with conformal field theory, fueled by the presence of a negative cosmological constant, facilitates the translation of complex quantum gravity problems into more manageable problems, giving rise to the emergence of the holographic principle \cite{Bousso:2002ju}.

In addition, black holes existing in anti-de Sitter  space can exhibit negative specific heat, from which there exists an energy loss that results in a decrease in temperature.

This intriguing feature gives rise to what is known as the Hawking-Page phase transition, during which black holes reach a state of thermal equilibrium with a thermal anti-de Sitter space \cite{Eom:2022nwc}.

As above stated, at asymptotic regime, however, we require $w$ will be independent of the coordinates.

The issue of having radial dependence is well-known for black holes surrounded by quintessence, as it happens for  the Kiselev solution \cite{Kiselev:2002dx}. This caveat  has been widely criticized recently, see e.g. \cite{Visser:2019brz}. Thus, it appears evident that a spherical symmetric solution gives rise to $r$-dependent pressure and density that cannot represent neither quintessence nor dark energy at all radii. Hence, a dark energy environment can be described \emph{only if} the radial coordinate does not appear in the equation of state parameter, pressure and density.

In support of this fact, we recall that in a homogeneous and isotropic universe, the cosmological principle imposes limits over the dark energy equation of state. Particularly, the equation of state parameter cannot be function of radial coordinates, but only of cosmic time. Analogously, whichever equation of state mimickers of dark energy has to show no radial dependence if we want to resemble the dark energy effects.

In the case of quintessence, in addition to the above requirement, we also need that the equation of state parameter is not phantom, namely it turns out to be larger than $-1$ but smaller than zero \cite{Copeland:2006wr,Tsujikawa:2013fta}.

For a metric under our form, this prerogative may occur at $r=0$ and asymptotically as $r\rightarrow\infty$.

\subsection{Reproducing quintessence}

Given the above, in general relativity, the weak energy condition (WEC) is given by
\begin{equation}\label{ECs0}
T_{\mu\nu}\phi^\mu \phi^\nu\geq0 \,
\end{equation}
where
$\phi^\mu$ is a timelike vector. According to the cosmological principle, WEC is equivalent to
\begin{equation}\label{ECs1}
\rho \geq 0, \quad \rho + P \geq 0,
\end{equation}
where $P=P_r=P_t$.

As mentioned in the introduction, the $\Lambda$CDM paradigm serves as the standard kinematic model for describing the large-scale dynamics, and is characterized by a de Sitter phase, where the energy density is equivalent to $\Lambda$. By virtue of the above conditions, it appears therefore  intuitive to employ

\begin{itemize}
\item[-] a regular spherically-symmetric spacetime  \cite{Bambi:2014nta,DuttaRoy:2022ytr,Malafarina:2022wmx,Bronnikov1979,Gonzalez-Diaz,Poisson,Dymnikova,Bronnikov:2006fu, Balart:2014cga, Fan:2016hvf, Toshmatov:2018cks, Toshmatov:2017zpr, Carballo-Rubio:2018pmi, Gibbons:1985ac, Bambi2013PhLB, Borde1994PhRvD,Barrabes,Bogojevic,Cabo,Hayward201,Jusufi2020,Ghosh2015EPJC,ToshmatovPhysRevD,Mustapha2014PhRvD,Heydari-Fard, 2023PhRvD.107h4048R,2022PhRvD.106d4031R,2023arXiv230317296L,2018PhRvD..97f4021T},  allowing for the inclusion of a vacuum energy component that can exhibit a de Sitter phase with a unspecified energy density, denoted by $\Lambda$;
\item[-] an equation of state parameter, $w$, asymptotically approaching a constant value, consistent with experimental constraints, as those derived from Planck observations \cite{Planck2018}, \textit{i.e.}, enabling the existence of dark energy, violating the WEC.
\end{itemize}

It is well-known that a scalar field, $\phi$, being non-dynamical and transporting vacuum energy under the form of a de Sitter phase, might undergo a \emph{slow-roll} regime \cite{Giambo2024}, in analogy to inflationary scenarios \cite{Linde:2007fr} and mimicking the role of $\Lambda$ in the $\Lambda$CDM model \cite{Capozziello:2019cav}.

In addition, to satisfy the aforementioned conditions, we can draw inspiration from the Hayward black hole \cite{Hayward:2005gi}, where the core exhibits a non-zero vacuum energy term, \textit{i.e.}, $1 - f(r,\phi) \propto r^2$ for $r\simeq 0$.

In our approach, we extend the Hayward solution: there, in fact, the metric tends to Schwarzschild at $r\gg1$, whereas in our case we require $1 - f(r,\phi) \propto r^2$ for $r\gg1 $ as well as for $r\simeq0$.

By virtue of the two possible deformations, that we described above, we propose  to recast Eq. \eqref{sferico2} plugging Eq. \eqref{secondadeformazione},

\begin{equation}
ds^2 =  -\left(1-\frac{\Lambda(\phi)}{3} r^{2+\epsilon(\phi)}\right) dt^2 +\frac{1}{g(r,\phi)} dr^2 + r^2 d \Omega^{2}\,,
\label{sfericofinale}
\end{equation}
where the concept of de Sitter-like solution arises in the $\epsilon=\epsilon(\phi)$ and $\Lambda=\Lambda(\phi)$ terms, having that,
 \begin{itemize}
     \item[-] $\Lambda$ is constant vacuum energy term with respect to $r$. Its sign is not specified \emph{a priori} and depends on $\phi$;
     \item[-] $\epsilon\rightarrow0$ as $\phi\rightarrow0$, namely without dark energy no  thermal bath surrounding the solution is expected;
     \item[-] $g(r)$ might be regular at $r=0$, diverging at $r\rightarrow\infty$ as $g\sim r^2$, \textit{i.e.}, restoring a de Sitter behavior.
 \end{itemize}

Thus, we evaluate the Einstein
equations adopting Eq. \eqref{sfericofinale}
\begin{subequations}
\begin{align}
{T_t}^t &=\frac{(r g(r))^\prime -1  }{r^2}\,, \label{eq:Ttt}\\[2mm]
{T_r}^r &=
\frac{r f'(r)+f(r)}{r^2 f(r)}g(r)-\frac{1}{r^2}=
\frac{1}{r^2}\left(\frac{1-(\Lambda/3)  (\epsilon +3) r^{\epsilon +2}}{1-(\Lambda/3)  r^{\epsilon +2}}g(r)-1\right)
  \label{eq:Trr} \\[2mm]
{T_\theta}^\theta &=
\frac{1}{4} \left(\frac{f'(r)}{f(r)}+\frac{2}{r}\right) g'(r)+\frac{2 f(r) \left(r f''(r)+f'(r)\right)-r f'(r)^2}{4 r f(r)^2}g(r) \notag\\
&=\frac{1}{2r}\cdot \frac{1-(\Lambda/6)  (\epsilon +4) r^{\epsilon +2}}{1-(\Lambda/3)  r^{\epsilon +2}}
g'(r)+
\frac{\Lambda  (\epsilon +2)^2 \left(1-(\Lambda/6)  r^{\epsilon +2}\right)}{6 \left(1-(\Lambda/3)  r^{\epsilon +2}\right)^2}r^{\epsilon },\label{eq:Tyy}
\end{align}
\end{subequations}

with ${T_\phi}^\phi  = {T_\theta}^\theta$, and the prime indicates the derivative with respect to the radial coordinate, $r$.

To guarantee isotropy on pressure, as in dark energy cases, we require $P_r=P_t=P$ as above stated, producing a linear ODE for the unknown $g(r)$ that can be integrated to give:

    \begin{equation}\label{eq:epsnon0}
g(r)= f(r)
\frac{ \, _2F_1\left(-\frac{2}{\epsilon +2},-\frac{\epsilon }{\epsilon +4};\frac{\epsilon }{\epsilon +2};\frac{\Lambda}{6} (\epsilon +4)r^{\epsilon +2} \right)+k_0 r^2}{\left(1-\frac{\Lambda}{6}  (\epsilon +4) r^{\epsilon +2}\right)^{\frac{\epsilon }{\epsilon +4}+1}}
\end{equation}

where $k_0$ is an integration constant, consequence of the first order differential equation to solve in order to address isotropy, whereas $_2F_1(a,b;c;z)$ is the analytical continuation of the hypergeometric function {\cite{Gasper_Rahman_2004}}.

We distinguish two cases, either $\epsilon\geq0$ or $\epsilon<0$, in which we see that for positive $\epsilon$ the functional behavior of the lapse function, $f(r)$, resembles a de Sitter-like form. In the simplest case, without the presence of an external field, we assume that the dark energy density is solely induced by the cosmological constant $\Lambda$.  The non-zero value of $\epsilon$, instead, is associated with the effects of the temperature becoming non-negligible.

\section{Varying $\Lambda(\phi)$ with $\epsilon=0$}\label{sezione3}

In order to guarantee that $g=1$ at $r=0$ and $g\sim r^2$, at $r\rightarrow\infty$, we first specialize on the simplest case  $\epsilon=0$, having

\begin{eqnarray}\label{eq:g11-eps0}
g(r)=\left(1-\tfrac\Lambda 3  r^2\right)\frac{1+k_0 r^2 }{1-2 \Lambda  r^2/3}\,,
\end{eqnarray}

where $k_0$ is an integration constant that will be physically reinterpreted later. Now, only notice that if $k_0=0$, $g$ converges at large radii, while diverges whether $k_0\neq0$. Hence, to address the de Sitter behavior at large radii, $g\sim r^2$, the natural recipe is to take $k_0\neq0$.

Particularly, the behavior at infinity immediately shows that a {genuine de Sitter} solution is recovered if
\begin{equation}
g\sim \frac{k_0\,r^2}{2}\rightarrow\,k_0\simeq -\frac23\Lambda\,,
\end{equation}
while  around $r=0$, up to the second order in $r$, we have

\begin{equation}\label{gasint}
g\sim 1+\left({k_0} + \frac{\Lambda}{3}\right) r^2+\ldots\,.
\end{equation}

Moreover,  the Ricci scalar reads
\begin{equation}\label{ricci}
R=-
\frac{6 k_0 \left(4 \Lambda ^2 r^4-11 \Lambda  r^2+9\right)+4 \Lambda ^2 r^2}{\left(3-2 \Lambda  r^2\right)^2},
\end{equation}

while the thermodynamic  quantities related to the energy momentum tensor, {i.e. the energy density $\rho$ and the pressure $P$,} yield

\begin{subequations}
\begin{align}
\rho &= \frac{\Lambda  \left(2 \Lambda  r^2-9\right)-3 k_0 \left(2 \Lambda ^2 r^4-7 \Lambda  r^2+9\right)}{\left(3-2 \Lambda  r^2\right)^2},
\label{eq:rho-eps0}\\
P &= \frac{3 k_0 \left(\Lambda  r^2-1\right)+\Lambda }{2 \Lambda  r^2-3}\,.\label{eq:P-eps0}
\end{align}
\end{subequations}

Clearly, we are forced to stress  that both $1-{\frac\Lambda{3}} r^2>0$ and $g(r)$, should have the same sign.
In view of \eqref{eq:g11-eps0} this means that
$$
\frac{ 1+k_0 r^2}{1-2 \Lambda  r^2/3}>0.
$$

This appears possible since  $\Lambda=\Lambda(\phi)$, and consequently its sign can be variable, depending on $\phi$.

{ At this stage, we can naturally provide a physical explanation of $\phi$.

The free parameter, $\Lambda(\phi)$ is a function of the environment field, $\phi$, only.

Consequently, the presence of $\phi$ can modify the sign of $\Lambda$ by only assuming that $\Lambda$ is analytical in terms of $\phi$. So, acting as an external temperature that does not explicitly depend on coordinates, to change sign, $\Lambda(\phi)$ can be parameterized by\footnote{{ This choice is constructed by supposing that $\Lambda$ is parameterized near a maximum of a phenomenological potential, say $\Lambda\simeq \Lambda_0-\Lambda_1\phi^2$, where $\Lambda_0$ matches the bare cosmological constant, whereas $\Lambda_1$ is intimately related to the phion mass associated with the field, $\phi$. Hence, rescaling the constants properly, we obtain the form in which explicitly the critical value, $\phi_c^2\equiv \frac{\Lambda_0}{\Lambda_1}$ appears. This scenario is inspired from the functional behavior of an hilltop potential in inflationary theory, albeit in this contexts it only represents a simple parametrization that enables us to switch from de Sitter to anti-de Sitter spacetimes.} }

\begin{equation}
\Lambda(\phi)=\Lambda_B\left(1-\frac{\phi^2}{\phi_c^2}\right)\,,
\end{equation}

that resembles similar approaches well-established in the context of symmetry breaking potentials \cite{Belfiglio:2023rxb,Siegel:1999ew,Luongo:2018lgy}.

Immediately, the physical meaning of $\phi$ appears that of a temperature whose effect consists in the change of shapes of the free parameters of our metric that, albeit constants with respect to coordinates, appear not to be constant in terms of the free temperature.

Precisely, at the critical temperature, $\phi_c$,  the transition occurs and $\Lambda$ vanishes, whereas before and after the transition we respectively have positive and negative vacuum energy defined.

The corresponding spacetime properties are thus associated with de Sitter and anti-de Sitter cases for the above cases, respectively.

Phrasing it differently, in order to guarantee that the effective  mass  changes sign, one assumes that there exists an external field, $\phi$, acting as temperature, able to induce a transition, at a given  critical value,  $\phi_c$.

}

Consequently, we can claim that

\begin{itemize}
    \item[-] $\phi_c$ is a \emph{black hole threshold}, since as we will focus later, the case $\Lambda>0$ is related to a singular solution,
    \item[-] the \emph{bare} cosmological constant, $\Lambda_B$ is a fixed, arbitrarily positive quantity,  that fixes physical dimensions. It is by definition $\Lambda(0)\equiv\Lambda_B>0$,
    \item[-] the transition of sign occurs passing from the configuration $\phi>\phi_c$ to the opposite, namely $\phi<\phi_c$,
    \item[-] since the field $\phi$ is non-dynamical, the corresponding effect to change the sign on $\Lambda$ does not alter the treatment but leaves open to two distinct cases, say $\Lambda$ positive or negative.
\end{itemize}

Consequently, two solutions are available from Eq. \eqref{ricci}, corresponding to regular and singular configurations,

\begin{subequations}\label{bhorbh}
\begin{align}
&\Lambda>0\Leftrightarrow {\rm singular\,solution,}\\
&\Lambda<0\Leftrightarrow {\rm regular\,solution.}
\end{align}
\end{subequations}

{ To argue these two behaviors, we can easily take into account the Ricci scalar, Eq. \eqref{ricci}. Immediately, one can see that it exhibits a singularity at $r=\sqrt{\frac{3}{2\Lambda}}$ for $\Lambda>0$. This singularity is obviously  eliminated if $\Lambda<0$. This property holds whichever values $k_0$ acquires.  Consequently, the solution appears de Sitter-like in the case of black holes only, while anti-de Sitter in the opposite case.

}

This implies two very different situations,  basically depending on the sign of $\Lambda$ and since $\phi_c$ is a threshold, in principle the solution can even pass from being regular to singular.

Below, we analyze in detail the two conditions on different signs for $\Lambda(\phi)$. Specifically, we here confuse $\Lambda$ with $\Lambda_B$, once its sign is imposed, \textit{i.e.}, as $\phi$ is fixed.

\subsection{Singular solution}

Assuming $\Lambda>0$, we cannot have the solution defined for all $r>0$, except for the deSitter case $k_0=-2\Lambda/3$. The following cases occur, depending on $k_0$.

\begin{enumerate}
\item{$k_0\le-2\Lambda/3$.}
The solution is defined in the set
\begin{equation}
r\in]0,1/\sqrt{-k_0}[\,\cup\, ]\sqrt{3/(2\Lambda)},+\infty[\,,
\end{equation}
and let us separately consider the models corresponding to the above two patches.

The first patch $r\in]0,1/\sqrt{-k_0}[$ is regular up to $r=0$ included, and does not have horizons. Energy density remains positive and the pressure is negative, but both are finite.

The second patch has a singular boundary at
\begin{equation}\label{eq:sing}
    r=\sqrt{\frac3{2\Lambda}},
\end{equation}
that disappears in the limit case $k_0=-2\Lambda/3$, where we obtain the static de Sitter patch defined for all $r>0$. Except this case, the singularity is present and contained in the stationary region bounded by the horizon
\begin{equation}\label{eq:hor1}
    r=\sqrt{\frac3{\Lambda}}.
\end{equation}
For greater values of $r$ is well defined and regular. Again, energy is positive and pressure is negative, and WEC is satisfied in the stationary region only, since $\rho+p$ change sign on the horizon \eqref{eq:hor1}.

\item{${-2\Lambda/3<k_0<-\Lambda/3}$.}
The solution is defined in the set
\begin{equation}
r\in]0,\sqrt{3/(2\Lambda)}[\,\cup\, ]1/\sqrt{-k_0},+\infty[\,,
\end{equation}
Again, let us analyze the two connected components separately.
The first is a solution with a regular centre and a singular non--central boundary at $r=\sqrt{3/(2\Lambda)}$ again, where the energy negatively diverges and the WEC does not hold.

The other patch has a horizon as \eqref{eq:hor1} but it is regular up to the boundary $r=1/\sqrt{-k_0}$, near which the energy becomes negative, which makes this solution eligible to build wormholes models, as done in Ref. \cite{Molina:2011mc}. WEC holds only outside the stationary region in this case.

\item{${-\Lambda/3\le k_0<0}$.} Similar to the above case, with the only exception that the outer patch $r\in]1/\sqrt{-k_0},+\infty[$ is not stationary ($\partial_t$ is spacelike), no horizons forms and WEC holds throughout.

\item{${k_0\ge 0}$.} In this case the outer patch of the previous case disappears,
and then only the patch defined in the right neighborhood of $r=0$ given by $
r\in]0, \sqrt{3/(2\Lambda)}[\,,
$ exists,
that again has a non--central singularity as in Eq. \eqref{eq:sing}. The energy density is always negative, whereas the pressure is positive in a neighborhood of $r=0$ when $k_0>\Lambda/3$, and then changes sign in the approach to the singularity. Here, WEC is never satisfied.
\end{enumerate}

\subsection{Regular solution}

Let us now take into account the case $\Lambda<0$, where consequently the $00$ component of the metric is well-defined $\forall r>0$. From Eq. \eqref{eq:g11-eps0}, taking $k_0<0$ results in a solution defined only for

\begin{equation}
 r<\frac{1}{\sqrt{-k_0}}.
\end{equation}
Therefore, from now on we will consider the case when $k_0\ge 0$, in such a way that $\forall r>0$ the solution is well-defined and, in view of Eqs. \eqref{eq:rho-eps0}--\eqref{eq:P-eps0}, also regular. Since $\Lambda<0$, no horizon forms here, see eqn \eqref{eq:hor1}.

However, in this case we cannot expect a positive energy density everywhere. In particular, when $k_0<-\Lambda/3$, then a right neighborhood of $r=0$ exists such that $\rho>0$, but then it becomes negative outside this neighborhood. When $k_0\ge -\Lambda/3$, instead, the energy will be negative on the whole solution. The pressure $P$ instead is  always positive regardless of the value of the (positive) constant $k_0$. Finally, the WEC is satisfied  when $k_0\le-{\tfrac{2\Lambda}3}$.

From Eqs. \eqref{eq:rho-eps0}--\eqref{eq:P-eps0} we have, asymptotically
\begin{align}\label{rhoasint}
&\rho\simeq (-3 k_0-\Lambda )-\frac{5}{9} r^2 \left(3 k_0 \Lambda +2 \Lambda ^2\right),\quad\text{as\ }{r\to 0^+},\\
&\rho\simeq-\frac{3}{2}k_0,\quad\text{as\ }r\to+\infty,
\end{align}
and
\begin{align}\label{pasint}
&P\simeq \frac{1}{3} (3 k_0-\Lambda )+\frac{1}{9} r^2 \left(-3 k_0 \Lambda -2 \Lambda ^2\right),\quad\text{as\ }{r\to 0^+},\\
&P\simeq \frac{3}{2}k_0,\quad\text{as\ }r\to+\infty,\label{pasint2}
\end{align}
implying again, besides $k_0>0$, even   $k_0\rightarrow\frac{2}{3}\Lambda$, to resemble a quintessence slow-roll regime. It is essential to emphasize that the value of $\Lambda$ is not predetermined \emph{a priori} and its constraints can be inferred from the Planck satellite findings \cite{Planck2018}.

The corresponding equation of state {parameter} behaves

\begin{equation*}
        w \simeq \begin{cases}
               \,\,\,\frac{1}{3}\left(\frac{\Lambda-3k_0}{\Lambda+3k_0}\right)\,
 \quad &\text{if\, $r\rightarrow0$,} \\
                        \,\\
                        \,-1\quad &\text{if\, $r\gg1$.}
                    \end{cases}
\end{equation*}

\begin{figure*}[h]
  \centering
    \includegraphics[width=.5\linewidth]{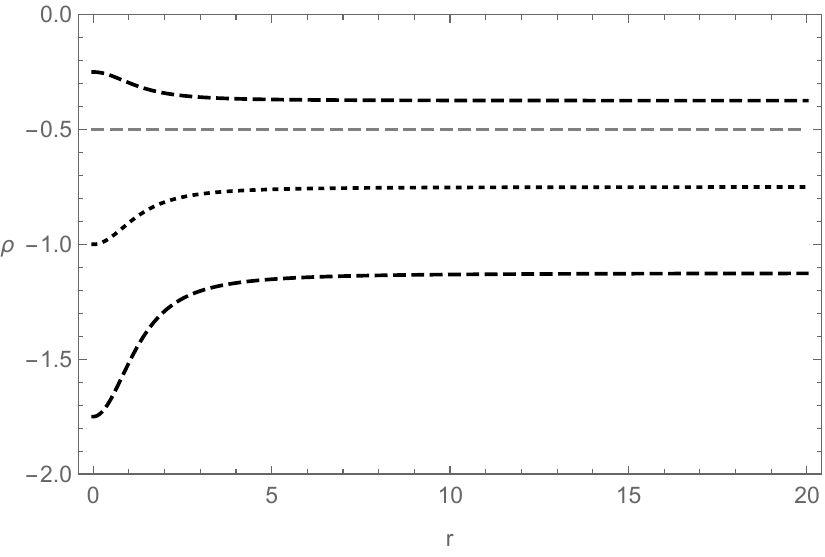}
  \hspace{0mm}\\
\includegraphics[width=.5\linewidth]{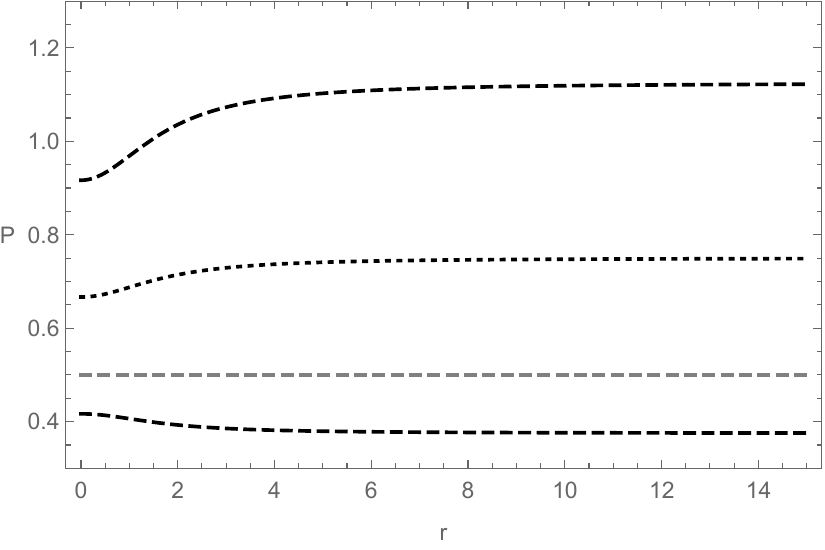}
\hspace{0.8mm}\\
    \includegraphics[width=.5\linewidth]{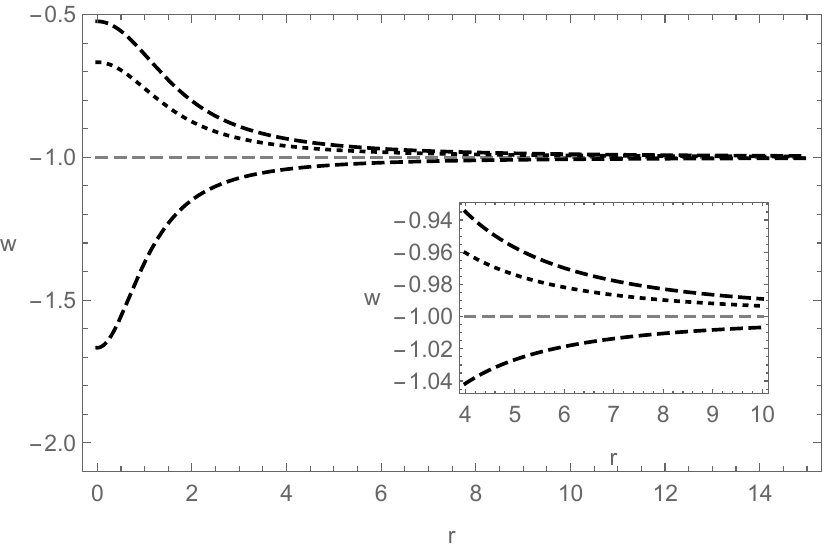}
    \vspace{7mm}\\
  \caption{Plots of thermodynamic quantities characterizing the fluid of our de Sitter-like solution, density (top), pressure (central), equation of state parameter (bottom).  The indicative value for the de Sitter phase is $\Lambda=-\frac{1}{2}$ and the different curves are associated with distinct sets of $k_0$. The gray dashed line yields the case $k_0=-\frac{2}{3}\Lambda$ that corresponds to the cosmological constant case. Here, $k_0=-\{0.5;1.0;1.5\}\times\Lambda$, specifically, the dashed black lines are associated to the upper and lower values of $k_0$, whereas the dotted line to $k_0=\Lambda$. The plot of $w$ provides convergence to the cosmological constant case, $w\simeq-1$ as $r\gg1$. A subplot shows the main (slight) differences of the various {equation of state parameters}, confirming that $P\rightarrow-\Lambda$ and $\rho\rightarrow\Lambda$, as $r\rightarrow\infty$.}
  \label{figura}
\end{figure*}

From the equation of state, supposedly, it seems that even the simplest solution $k_0=0$ can be employed, revealing that as $r$ approaches zero, the equation of state {parameter} tends to $w=\frac{1}{3}$, which indicates a fluid similar to radiation. However, as we progress in the text, we demonstrate that this solution is not purely radiation. We accomplish this by analyzing the sound speed and reinterpreting our fluid, demonstrating that at small radii, the only plausible interpretation is stiff matter.  We display in Figs. \ref{figura} the behaviors of $\rho, P$ and $w$ with indicative values of $\Lambda$.

\subsection{An alternative derivation}

Interestingly, one can obtain the same solution that we evaluated invoking a spherically symmetric singular solution modified by an additive deformation  \cite{Molina:2011mc}. Let us impose isotropy   a static metric in the form of Eq. \eqref{sferico0} and  assume that the metric coefficients are written in a $\delta$--expansion form, where $\delta$ measures the deviation from vacuum solution with cosmological constant, \textit{i.e.}, from the Schwarzschild--de Sitter spacetime. Thus, we have
\begin{align}
    f(r)&=\chi(r)+\delta f_1(r)+\ldots,\\
    g(r)&=\chi(r)+\delta g_1(r)+\ldots,
\end{align}
where
\begin{equation}
    \chi(r)=1-\frac{2M}{r}-\frac{\Lambda}{3}r^2.
\end{equation}
The isotropy condition can be extremely simplified with the additional ansatz
\begin{equation}
    f_1(r)=\chi(r)\alpha(r),\qquad g_1(r)=\chi(r)\beta(r).
\end{equation}
Moreover, working out the \textit{exact linear solutions}, inferred as $\alpha(r)=0$:

\begin{equation}
f(r)=\chi(r),\qquad g(r)=\chi(r)(1+\delta\beta(r)),
\end{equation}
we get the isotropy condition to become
\begin{equation}\label{eq:ODEb}
   \frac{\beta '(r)}{\beta (r)}= -\frac{6}{3 M+2 \Lambda  r^3-3 r}.
\end{equation}
In such a case, the energy momentum tensor turns out to be exactly linear in $\delta$, say
\begin{align}
    T^0_0&=-\Lambda+\left(-\Lambda +\frac{2}{r^2}-\frac{3 (r-3 M)}{r^2 \left(3 M+2 \Lambda  r^3-3 r\right)}\right)\,\delta\beta(r), \\
    T^i_i&=-\Lambda+\left(-\Lambda+\frac{1}{r^2} \right)\delta  \beta (r),
\end{align}
    where $\beta(r)$ solves Eq. \eqref{eq:ODEb}, manifesting as a linear differential equation. Also note that the $\Lambda$ term above is embodied in $T^\mu_\nu$ because we are considering field equations in the form \eqref{einstein_equations}.  This would allow in principle to absorb $\delta$ in the  constant (say, $D$) coming from integrating Eq. \eqref{eq:ODEb}, setting $C:=D\cdot\delta$ and considering the family of solutions parameterized by the only real parameter $C$.

In principle $f(r)$ and $g(r)$ thus obtained
have not necessarily the same sign $\forall r>0$, which is a requirement to address for any Lorentzian geometry, as previously stated.

Moreover, as $M>0$ and $\Lambda<1/9M^2$, then $f(r)$ and $g(r)$ are positive in a neighborhood $(r_+,r_c)$, where $r_+$ is the black hole horizon and $r_c$ is the cosmological horizon of the geometry.
In addition, as $C>0$, one can show the existence of a value $r_0\in(r_+,r_c)$ where curvature diverges and the static geometry is then well-defined within the interval $(r_0,r_c)$, possessing a naked singularity when $r\to r_0$.

When $C<0$ the geometry is more complex because $g(r)$ has a zero $r_{thr}\in(r_+,r_c)$, although the metric is regular there.

If the gravitational charge vanishes, \textit{i.e.}, $M=0$, we recover Eq. \eqref{eq:g11-eps0}, with $f=1-\frac{\Lambda}{3}r^2$, namely we reobtain our solution when $\epsilon=0$.

It is also possible to demonstrate that, due to the assumptions on $\Lambda$ and $C$, we have that the metric is defined  $\forall r\in(r_{thr},r_c)$ and {by attaching two such solutions at $r=r_{thr}$, a wormhole solution can be obtained}, see Ref.  \cite{Molina:2011mc}.

In summary, {metric \eqref{sferico0} is capable of reproducing a singular and regular solutions. In case of additional requirements it is also possible to get wormholes}.

It is therefore natural to compare our solution with the most prominent ones transporting vacuum energy, say the de Sitter and Hayward solutions, to check the main differences with our finding.

\subsection{Comparing our metric with the Hayward and  de Sitter solutions}

In this section, we compare our findings with the well-established Hayward and de Sitter spacetimes. Specifically, the lapse function $f(r)$ for the Hayward black hole is
\begin{equation}
f (r) =  g(r) = 1 - \frac{2 M r^2}{r^3 + 2 a^2}\,,
\end{equation}
where $M$ is the mass of the black hole, $r$ is the radial coordinate and $a$ is a constant. Easily, let us assume
\begin{equation}
a=\sqrt{\frac{3}{2\Lambda}},
\end{equation}
in order to recover the relevant property that this metric transports vacuum energy, in analogy to the pure de Sitter solution.

As a consequence, our comparison can be mainly focused  around $r\simeq0$, since the asymptotic case, $r\rightarrow\infty$, tends to Schwarzschild at large radii for the Hayward spacetime.

The main differences are listed below.

\begin{itemize}
    \item[-] The Hayward solution is not isotropic. It falls into the class of \textit{anisotropic generalizations} of de Sitter spacetime discussed in Ref. \cite{Giambo:2002wr}. The energy momentum tensor is written
involving
\begin{align}
    \rho_H&= \frac{18 M \Lambda}{(3 + r^3 \Lambda)^2},\\
P_{r,H}&=-\rho_H,\\
P_{t,H}&=P_r+\frac{54 M r^3 \Lambda^2}{(3 + r^3 \Lambda)^3},
\end{align}
where the subscript $H$ clearly refers to as ``Hayward".
\item[-] At small radii, up to fourth order, we have
\begin{align}
    \rho_H&\simeq 2M\Lambda-\frac{4}{3}M\Lambda^2r^3+\mathcal O(r^4)\\
    P_{t,H}&=-2M\Lambda +\frac{10}{3}M\Lambda^2 r^3+\mathcal O(r^4)\,.
\end{align}

\item[-] The speed of perturbations is due to the sound speed, say
\begin{equation}\label{suono}
c_s^2=\frac{\partial P}{\partial\rho}\,,
\end{equation}
and then, since $P=P(r)$ and $\rho=\rho(r)$, it is easy to show that
\begin{equation}\label{eq:cs2-def}
c_s^2\cong \frac{P'(r)}{\rho'(r)}\,,
\end{equation}
giving for the Hayward spacetime
\begin{align}
    c_{s,r}^2&=-1\\
    c_{s,t}^2&=2-\frac{27}{6+2\Lambda r^3}\,,
\end{align}
that in both cases appears unphysical, because negative-definite.

Remarkably, it is needful to stress that the sound speed represents the velocity of perturbations of a given fluid. In the case of negative {equation of state parameters}, it can happen that the squared sound speed turns out to be negative, producing instabilities. Clearly, not all the negative {equation of state parameters} are unstable, e.g., the $\Lambda$CDM paradigm presents an equation of state {parameter} identically equal to $-1$, albeit the sound speed is identically zero, as well as in the dark fluid case. In quintessence realm, the sound speed is identically $1$ and, again, positive definite, producing stiff matter, i.e., the perturbations evolve with the speed of light.

\begin{figure*}[h]
  \centering
    \includegraphics[width=.8\linewidth]{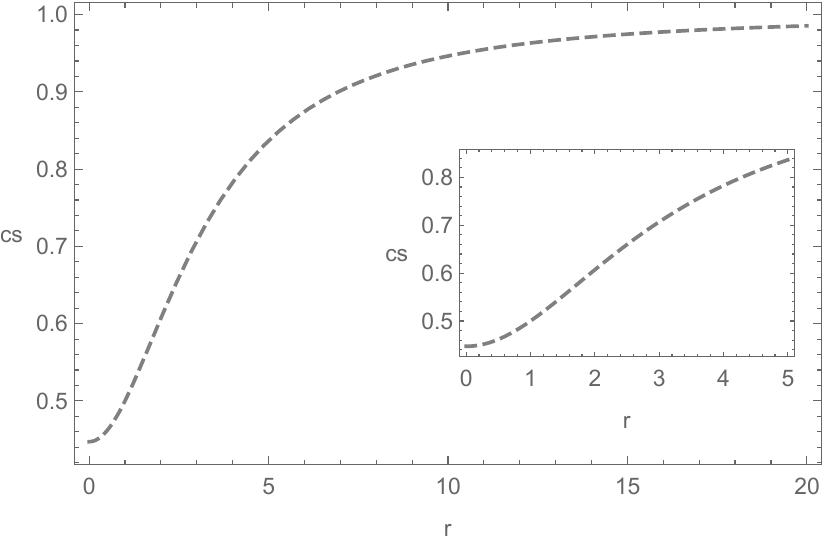}
  \hspace{0mm}\\
  \caption{Sound speed of our de Sitter-like model with fixed $\Lambda=-0.5$. The sound speed is always positive definite, as in Eq. \eqref{eq:cs2}. It approaches stiff matter behavior at large radii, indicating quintessence with a negative-definite potential, associated with exotic matter. Violating isotropy implies unstable sound speeds, see Eqs. \eqref{csrun1} - \eqref{cstun1}.}
  \label{sonido}
\end{figure*}

Nevertheless, for our model we compute
\begin{equation}\label{eq:cs2}
c_s^2\cong \frac{P'(r)}{\rho'(r)}=
\frac{3-2 \Lambda  r^2}{15-2 \Lambda  r^2}
\,,
\end{equation}
non-depending on $k_0$.

{ Given $\Lambda$, the assumption of isotropy generates a \emph{one-parameter family of metrics}, depending on the integration constant $k_0$. Specifying $k_0$ would amount to specifying a particular member of the family. Accordingly, the assumption of isotropy completely determines $c_s^2$, without the need for further assumptions, since $k_0$ represents a shift in the vacuum energy carried on by the metric. To better see this point, it could be interesting to compute the Misner-Sharp mass, \textit{i.e.}, as a quasi-local mass contained within a sphere of radius $r$. It reads,

\begin{equation}
m_{M-S}(r)=\frac{ \Lambda + k_0 (3 - r^2 \Lambda)}{(1 +    k_0 r^2) (3 - r^2 \Lambda)}r^3,
\end{equation}

The corresponding density, that in spherical symmetry is proportional to the Misner-Sharp mass over $r^3$, is at leading order around $r=0$,

\begin{equation}
m_{M-S}/r^3=  k_0 + \frac{\Lambda}{3} +\mathcal{O}(5)\,,
\end{equation}
showing that it turns out to be constant in the center of our model.

Accordingly, since the above constant is a combination of the two free parameters, $k_0$ acts as a rescaling constant over the vacuum energy term, $\Lambda$, as we clamed above.

Following the same interpretation, since the adiabatic sound speed is a ratio between two perturbations, the pressure perturbation in the numerator over the density perturbations, we do expect that it could not depend on $k_0$.

Physically, this implies that the fluid  perturbations are not influenced by the \emph{offset} imposed by the integration constant that determines the class of metrics, as naively expected. Further, in agreement with the above description, related to the Misner-Sharp mass, this property manifests into the degeneration between the two constants, $k_0$ and $\Lambda$ that can be, therefore, be chosen arbitrarily.\footnote{{ For the sake of clearness, we notice that the Ricci scalar vanishes at $r=0$ if $k_0=0$. This implies that, albeit $k_0$ is fully arbitrary and degenerate with $\Lambda$, its vanishing determines a corresponding vacuum solution. Moreover, the corresponding sign, if $k_0\neq0$ and $r=0$, implies to recover a de Sitter or anti-de Sitter Ricci scalar, respectively.}}

Hence, the functional form of the sound speed is positive-definite and leads to a precise fluid definition, as we will clarify later. The behavior of sound speed is therefore well-defined and prompted in Fig. \ref{sonido}.
}

\end{itemize}

In the de Sitter case, the density and pressure appear constant and equal to $\rho_{dS}=-P_{dS}=\Lambda$, with vanishing sound speed, in close analogy to the $\Lambda$CDM model. Moreover, contrary to our solution and to the de Sitter spacetime, the Hayward metric is manifestly constructed for accounting positive de Sitter phase, say $\Lambda>0$, exhibiting a divergence when $\Lambda<0$, appearing exactly the opposite of our case.

We will discuss the role played by the sound speed later, when we will find a direct interpretation of our fluid at small and large radial distances.

\section{Perturbing the metric: Varying $\Lambda(\phi)$ with  $\epsilon\neq0$}\label{sezione4}

In view of the above results, we inferred that the fluid associated with our solution transports vacuum energy and significantly differs from the de Sitter and Hayward solutions. It is therefore licit to expect that possible deformations can be, in principle, responsible for departures in our findings as we initially conjectured.

In this section, we show that the aforementioned deformations cannot provide deep changes in the corresponding physics. Specifically, we are interested in small deviations from the de Sitter-like case, therefore $\epsilon$ {can be taken negligibly small}.

Recalling all the above, one can state that we search for a bound on $\epsilon$ that implies a lapse function $f(r,\epsilon(\phi))$ with the property $f(r,0)=r^2$, as $\epsilon(0)=0$. Thus, if $f(r)=1-\tfrac\Lambda 3 r^{2+\epsilon}$ one finds the general solution given by Eq. \eqref{eq:epsnon0}.

There, we observe that
to have a Lorentzian metric then the fraction at the right-hand side of  \eqref{eq:epsnon0} must be positive and this results, at least when $|\epsilon|\ll 1$, in the condition $\Lambda<0$ (as in the $\epsilon=0$ case) and $k_0\ge k_{\text{crit}}$, where $k_{\text{crit}}$ is the positive constant given by
\begin{equation}
k_{\text{crit}}= -\frac{\Gamma \left(\frac{\epsilon }{\epsilon +2}\right) \Gamma \left(\frac{2}{\epsilon +2}-\frac{\epsilon }{\epsilon +4}\right)}{\Gamma \left(-\frac{\epsilon }{\epsilon +4}\right)}\left[-\frac\Lambda 6 (\epsilon+4)\right]^{\frac{2}{\epsilon +2}},
\end{equation}
where $\Gamma(z)$ is the Euler Gamma function.

Noticeably, the condition $k_0\ge 0$ is not recovered in the limit $\epsilon\to 0$, since $\lim_{\epsilon\to 0}k_{\text{crit}}=-\tfrac{\Lambda}{3}$. {This  result can be interpreted as an instability feature of  the case $\epsilon\neq0$}. Clearly, $k_{{\rm crit}}$ is associated to the above $\phi_c$, \textit{i.e.}, it is intimately related to how we select the critical value of $\phi$, inducing the sign transition.

Quite interestingly, we remark that

\begin{subequations}
    \begin{align}
&\lim_{r\to 0^+}\rho(r)=-3k_0,\\
&\lim_{r\to 0^+}P(r)=k_0
    \end{align}
\end{subequations}

being independent of $\Lambda$ and fulfilling the fact that one cannot recover the case $k_0=0$ as $\epsilon\rightarrow0$.

We conclude that the corresponding solution is unstable and appears clearly unphysical in the description of a metric that transports vacuum energy.

\section{Interpreting the fluid}\label{sezione5}

The functional form of our equation of state easily suggests that by varying $k_0$ it is possible to infer different species of barotropic fluids described by our metric. For example, handling $k_0=0$, we discussed previously that it is possible to immediately see that the equation of state {parameter} reduces to $w=\frac{1}{3}$ in the case $\epsilon=0$. At a first glance, it can be possible to conclude that such a fluid mimes radiation, albeit from a more detailed analysis the conclusion appears different.

In addition, our model, with $\epsilon\neq0$ or $\epsilon=0$, shows that both pressure and energy tend to a nonzero constant ar $r\rightarrow0$, see Eqs. \eqref{rhoasint}--\eqref{pasint}.%:

Then, if one would restore the coincidence between $w$ and $c_s^2$, \textit{i.e.}, assuming a constant equation of state {parameter} around $r\simeq0$, then this would force the condition $k_0=\Lambda/12$, that is not acceptable if we want the exact profile given by Eq. \eqref{eq:g11-eps0} for every $r>0$.

Hence, phrasing it differently,  it is not possible to have $c_s^2=\omega$, near the center as one finds for barotropic dark energy models with constant equation of state {parameter}. This fact cannot occur since it would imply $k=\Lambda/12$, as stated, but, requiring a physical solution over $r\in[0,+\infty)$ it behooves us $\Lambda<0$ and $k\ge 0$.

So, one may explore different solutions such that $c_s^2=\omega$ as $r\to 0^+$, extending our cases. As a consequence, we may impose that both $\rho(r)$ and $P(r)$ vanish at the center, otherwise we would fall in the same drawback discussed above.
This further hypothesis will result into a new condition to impose on the metric {that we can introduce still keeping the isotropy condition on the pressures, $T^r_r=T^\theta_\theta$. Assuming a sufficient degree of regularity of the metric coefficients at the centre, the above conditions will produce a sequence of relations between the McLaurin coefficients of $f(r)$ and $g(r)$. In particular,
the lowest order free coefficient will turn out to be $f(0)$ -- that by local Cartesianity we can set to 1 -- and $f^{(3)}(0)$, that will be set equal to $6\alpha$ -- with $\alpha$ generic constant to fix with $\Lambda$ --}
having at leading order
\begin{subequations}
    \begin{align}
f(r)&=1+\alpha  r^3+ o(r^3),\label{alter1}        \\
g(r)&=1-3\alpha r^3+o(r^3),\label{alter2}
    \end{align}
\end{subequations}
and
\begin{subequations}
    \begin{align}
\rho(r)&=12\alpha r+o(r),\\
P(r)&=-\frac{9\alpha^2}{2}r^4+o(r^4).
    \end{align}
\end{subequations}

Consequently, if this further condition is fulfilled, the initial model can be adapted by updating it through  $f(r)=1+{\alpha r^3}$, with no higher order terms, just near the center. In our scenario, comparing our lapse function in the case $\epsilon\neq0$ and Eq. \eqref{alter1}, we see that it would correspond to $\epsilon=1$ and $\alpha\equiv -{\frac\Lambda{3}}$. This clearly violates the condition $\epsilon\ll1$. This new metric does not exhibit a manifest limit to the de Sitter spacetime. Hence, even by setting up the free coefficients, the metric cannot provide a $\Lambda$-plateau at $r=0$, being physically different than our proposed metric.

Thus, our spacetime appears easier and cannot reduce to a barotropic fluid with constant equation of state {parameter} at the center. A physical interpretation of our metric can be inferred by looking at the corresponding fluid. Thus, let us recall Eq. \eqref{eq:cs2}, computed at $\epsilon=0$,
\begin{subequations}
\begin{align}
c_s(0)&=\frac{1}{\sqrt{5}}\label{limitcs0}\,,\\
c_s(\infty)&=1\label{limitcsinfty}\,,
\end{align}
\end{subequations}
showing, as stated above, that $k_0$ does not play a significant role in determining the fluid. Noticing this, one concludes that the physics of our solution cannot be influenced by $k_0$ and therefore the radiation that apparently arose as $k_0=0$ cannot be seen as a suitable physical outcome.

On the other hand,  from Eqs. \eqref{limitcs0}--\eqref{limitcsinfty}, at infinite distances a stiff matter value of $c_s$ is found, but with an asymptotic equation of state {parameter},  $w\rightarrow-1$.

Immediately, one can notice that our scalar field, $\phi$, entering the metric as a temperature environment, can generically written by
\begin{subequations}
    \begin{align}
\rho&=X+V(\phi)\,,\label{barro1}\\
P&=X-V(\phi)\,,\label{barro2}
\end{align}
    \end{subequations}
where $X\equiv \frac{1}{2}g^{\mu\nu}\partial_\mu\phi\partial_\nu\phi$ with $V(\phi)$ the quoted generic scalar field potential that induces the effects of $\phi$.

Thus, from Eq. \eqref{suono}, we easily obtain
\begin{equation}
c_s^2=\frac{dP/dX}{d\rho/dX}=1\,,
\end{equation}
suggesting that our fluid mimes the sound speed of a scalar field with negative-definite potential, in the slow-roll regime, characterized by $|V(\phi)|\gg X$. In fact, as claimed above, $w=-1$ as $r\rightarrow\infty$.

This interpretation perfectly fits with our demand to require non-dynamical $\phi$ in order to induce a phase transition over the sign of $\Lambda$.

For the sake of completeness, another possibility could be  provided by quasi-quintessence fluid \cite{2010PhRvD..81d3520G,Chimento:2007da,Chimento:2010vu,Luongo:2018lgy,2022CQGra..39s5014D,Belfiglio:2023rxb} but there, although the same slow-roll regime can be achieved, the fluid provides a vanishing sound speed because the pressure is not function of $X$ and, so, this fluid cannot be recovered with our solution.

\subsection{Alternative conditions on the pressure}

In the context of  dark energy scenarios, isotropy has been previously considered. However, there is no compelling reason to separately examine cases where dark energy contributes only one single pressure while the other pressure vanishes or, more generally, to investigate the anisotropic case.

Examples of anisotropic dark energy models have been explored, for instance, through Bianchi metrics and acquired recently more emphasis in view of current developments that seem to indicate, although very slightly, departures from the genuine cosmological principle, see e.g. \cite{Luongo:2021nqh,Cao:2023eja,Krishnan:2022qbv,Shamir:2022wlx}. Clearly, quintessence is isotropic, so demanding to violate isotropy cannot work to get a quintessence asymptotic domain.

Assuming that the isotropy condition is violated and that $P_t=0$ with non-vanishing radial pressure \cite{Magli:1997qf}, in the case $\epsilon=0$, we obtain
\begin{equation}
g(r)=k_1\frac{3-\Lambda  r^2}{\left(3-2 \Lambda  r^2\right)^{3/2}},
\end{equation}
that furnishes
\begin{subequations}
    \begin{align}
\rho&=\frac{1}{r^2}\left[{1-3 \sqrt{3} k_1\frac{ \Lambda  r^2+3}{\left(3-2 \Lambda  r^2\right)^{5/2}}}\right],\\
P_r&=-\frac{1}{r^2}\left[{1+3 \sqrt{3} k_1\frac{ \Lambda  r^2-1}{\left(3-2 \Lambda  r^2\right)^{3/2}}}\right],
\end{align}
\end{subequations}

with $k_1$ an integration constant to be fixed.

On the other side, for $P_r=0$ and $P_t\neq0$, we have
\begin{equation}
g(r)=\frac{3 - \Lambda r^2 }{3\left(1 -  \Lambda r^2 \right)},
\end{equation}
that yield
\begin{subequations}
\begin{align}
\rho&=\frac{2\Lambda}{3}\frac{\left(\Lambda  r^2-3\right)}{ \left(\Lambda  r^2-1\right)^2},\\
P_t&=\frac{\Lambda^2}{3}\frac{r^2}{ \left(\Lambda  r^2-1\right)^2}.
\end{align}
\end{subequations}

Notice that the solution that involves only tangential pressure depends on less constants than the cases of isotropy and radial pressure only, as a consequence of the fact that the equation from which one argues $g(r)$ is not differential.

Indeed, it is evident that the scenario with tangential pressure alone is not favored. For any value of $\Lambda$, the pressure is always positive, but when considering negative values of $\Lambda$, the density becomes negative, suggesting the presence of exotic matter exclusively. The natural sign of $\Lambda$ is again negative, in order to avoid singularities. Accordingly, we infer
\begin{subequations}

    \begin{align}
        &c_{s,r}^2(0)=c_{s,r}^2(\infty)=-1\,,\label{csrun1}\\
        &c_{s,t}^2(0)=-\frac{1}{10},\quad c_{s,t}^2(\infty)=\frac{1}{2}\,,\label{cstun1}
    \end{align}
\end{subequations}

remarking that the models provide instabilities, since the sound speed is mainly negative. The limiting cases for the {equation of state parameters} yield
\begin{subequations}
    \begin{align}
        &w_{r}(\infty)=w_{r}(0)=-1\,,\\
        &w_{t}(\infty)=\frac{1}{2},\quad w_{t}(0)=0\,,
    \end{align}
\end{subequations}
showing that asymptotically the radial model approaches a cosmological constant, albeit the sound speeds are negative, \textit{i.e.}, unstable, while the tangential model provides a matter fluid-like, within the Zeldovich limit.

\begin{table*}
\footnotesize
\setlength{\tabcolsep}{0.5em}
\renewcommand{\arraystretch}{2}
\begin{tabular}{lcccccccccc}
\hline
\hline
$Model$ &  $Characteristic$ & $\Lambda<0$ & $\Lambda>0$    &   $w_r(\infty)$ & $w_t(\infty)$ &      $ Fluid-like$    \\

\hline
${\rm dS}$           &     ${\rm Isotropic}$
                & ${\rm Regular}$
                & ${\rm Regular}$ & ${\rm -1}$ & $\equiv w_r(\infty)$  & ${\rm CC}$
                \\
${\rm Hayward}$
                &     ${\rm Anisotropic}$
                & ${\rm Singular}$
                & ${\rm Regular}$ &  ${\rm -1}$ & $\infty$  & ${\rm -}$
                \\
${\rm  dS -\,like}$       &  ${\rm Isotropic}$ &  ${\rm Regular}$
                & ${\rm Singular}$
                & $-1$ & $\equiv w_r(\infty)$  & ${\rm Q}$
                \\
${\rm dS_r-like}$                &     ${\rm Anisotropic}$
                & ${\rm Regular}$
                & ${\rm Singular}$  & $-1$ & $-$ & $-$
                 \\
${\rm  dS_t-like}$
                &     ${\rm Anisotropic}$
                & ${\rm Regular}$
                & ${\rm Singular}$ & $-$ & $\frac{1}{2}$  & ${\rm Matter}$
                \\
\hline
\hline
\end{tabular}
\caption{Schematic summary of our solutions in comparison with other spacetimes. In particular, we consider our de Sitter-like metric, named ${\rm dS-like}$, compared with the de Sitter solution, ${\rm dS}$, the Hayward spacetime and with the alternative metrics obtained departing from the isotropic condition, namely ${\rm dS_r-like}$ and ${\rm dS_t-like}$ line elements, respectively the one with non-vanishing radial and tangential pressure. The fluid-like behavior stands for ${\rm Q}$ indicating \emph{quintessence} as it exhibits a negative effective potential in a slow roll regime, for ${\rm CC}$, indicating a dS phase that resembles the cosmological constant and finally ${\rm matter}$ that shows tangential matter-like fluid. The \emph{unspecified} results are indicated through ${\rm -}$, \textit{i.e.}, when either the fluid is not among the above classes or the result cannot expected. Among all the above scheme, it appears evident that the de Sitter solution is the unique transporting a pure cosmological constant. Our solution appears, on the other hand, better behaving than the Hayward spacetime, but with the great disadvantage of providing an anti-de Sitter phase. Departing from isotropy seems to result into misleading fluid-like asymptotic constituents. }
\label{tabella}
\end{table*}

\begin{figure*}[h]
  \centering
\hspace{1.mm}\\
    \includegraphics[width=.5\linewidth]{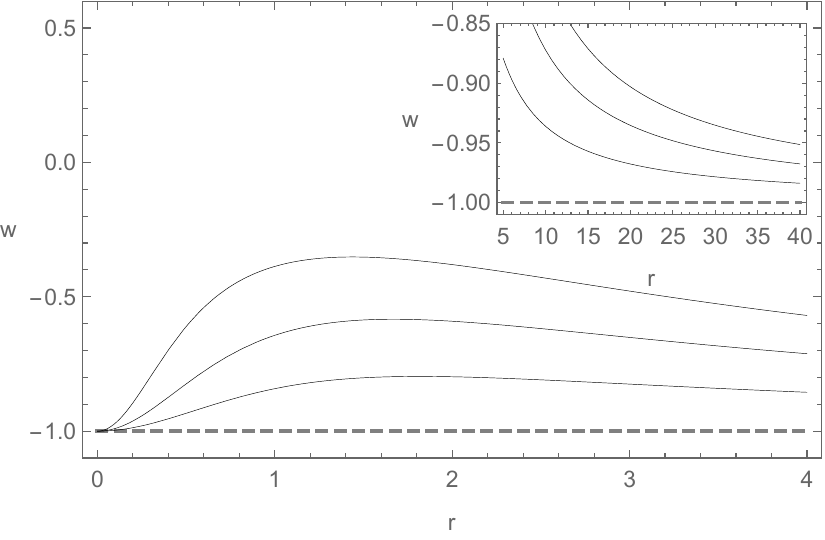}
  \hspace{0mm}\\
    \includegraphics[width=.5\linewidth]{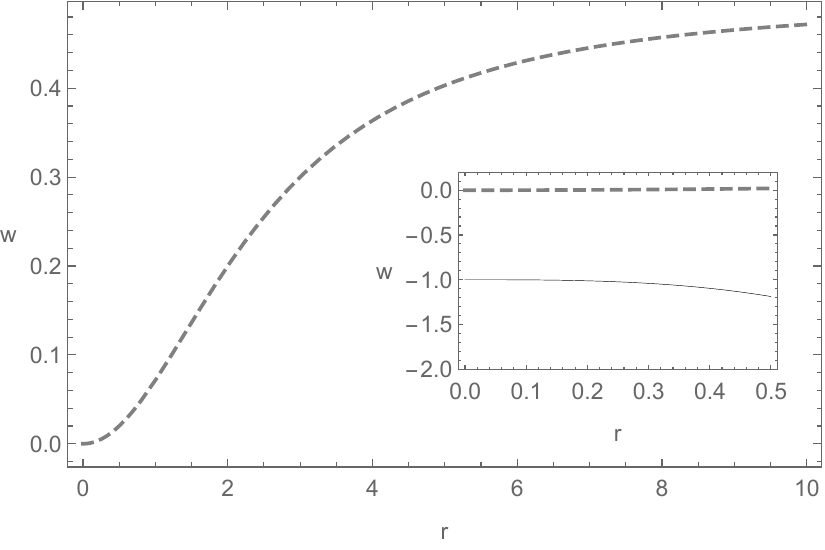}\hspace{0.8mm}\\
    \includegraphics[width=.5\linewidth]{w.pdf}
    \vspace{7mm}\\
  \caption{Plots showing the behaviors of {equation of state parameters}, with indicative value, $\Lambda=-0.5$. Notice that the radial equation of state depends on a further parameter, $k_1$. (Top) Radial equation of state with $k_1=-\{0.5;1.0;1.5\}\times\Lambda$. The subfigure focuses on differences at larger radii. (Central) Tangential equation of state with fixed $\Lambda$. The subfigure shows the difference with the Hayward equation of state. (Bottom) Plot of the equation of state {parameter}.}
  \label{figura2}
\end{figure*}

The natural behaviors of such equations of state are reported in Figs. \ref{figura2}, whereas all the here-developed cases are summarized in Tab. \ref{tabella}.

\section{Final outlooks}\label{sezione6}

In this paper, we investigated a class of spherically-symmetric solutions carrying vacuum energy in the form of a de Sitter-like phase represented by an effective cosmological constant. To achieve this, we left unspecified the sign of $\Lambda$ and considered the possibility that it can be modified through the existence of an external scalar field, denoted by $\phi$, which is non-dynamical. We assumed that this scalar field describes the temperature arising from the presence of a dark energy environment surrounding our spacetime solution. Consequently, we examined the implications of this approach at all radii.

Specifically, to investigate the phase transition, we considered two types of deformations on the metric. The first deformation acts on $\Lambda$, altering its sign, while the second modifies the functional form of $\sim r^2$, which is responsible for transporting vacuum energy. For the first transition, we introduced a toy model that allows for a change in the sign of $\Lambda$ once the external field is fixed. Consequently, we proposed the existence of a critical value of $\phi$, denoted as $\phi_c$, and found that the metric can pass through a regular to a singular configuration as this transition occurs.

In particular, we got our results by conventionally
considering the $00$-component of the metric, which exhibits behavior similar to a de Sitter solution, while imposing the condition of isotropic pressure. We justified the assumption of isotropy to ensure that the external dark energy, responsible for the presence of a temperature field, matches  the key characteristics measured by current observations. Our class of solutions is found to be regular or singular, depending on whether the cosmological constant is negative or positive, respectively. Consequently, we obtained both anti-de Sitter and de Sitter-like behaviors by making different choices for $\Lambda$.

Further, we slightly summarized some conditions to obtain wormhole solutions by considering  different additive deformations on the metric. In so doing, we demonstrated that more complicated deformations, different from additive,  appear unstable and physically-unmotivated.

To show this, we investigated the nature of the fluid and we found that it cannot be at the same time barotropic and providing $c_s^2=w$, with constant $w$, near the center. Hence, we investigated a further hypothesis on the metric in order to fulfill $c_s^2=w$ around the center. To this end, we demonstrated that this finding induces a modification that does not enable to transport vacuum energy in analogy to the de Sitter spacetime, violating the condition $\epsilon\ll1$.

Thus, we reinterpreted our fluid by virtue of a slow-roll quintessence by investigating the behaviors of the sound speed and equation of state at asymptotic regimes. Indeed, we showed that our solution provides stiff matter with $w=-1$ as $r\gg1$, confirming that the field dynamics is negligible within a slow-roll regime. Consequently, we concluded that our outcome appears as an \emph{isotropic de Sitter-like solution} that asymptotically acts as a quintessence fluid.

We also investigated the possible implications of violating isotropy, \textit{i.e.}, by having either non-zero radial pressure or non-zero tangential pressure only. Comparisons with the de Sitter and Hayward solutions have been also discussed, showing advantages and disadvantages of our findings.

{ Summarizing, our spacetime shows the following properties.

\begin{itemize}
    \item[-] It mimes quintessence at asymptotic regime and it is singular or regular depending on the the sign of $\Lambda$, differently from Hayward and de Sitter metrics. Moreover, in the case of ${\rm dS}_t$-like solution, the corresponding interpretation is purely matter, whereas for the dS model the cosmological constant behavior is recovered.
    \item[-] It transports vacuum energy, in analogy to Hayward and de Sitter metrics, but it exhibits isotropic pressures, differently from Hayward,
    \item[-] It naturally extends de Sitter, representing the simplest extension to it, relaxing the hypothesis of Schwarzschild coordinates.
    \item[-] It provides stable sound speed fluid perturbations, well-defined for given sets of free constants.

\end{itemize}

}

In view of the above, as  perspectives of our work, we intend to study alternatives that may carry on vacuum energy. Moreover, we want to work out metrics that can exhibit barotropic dark energy environment without passing through the existence of quintessence fields. Finally, breaking the spherical symmetry will also be investigated in future efforts.

\section*{Acknowledgements}
The work of OL is  partially financed by the Ministry of Education and Science of the Republic of Kazakhstan, Grant: IRN AP19680128.

\section*{References}
\bibliographystyle{ieeetr}

\end{document}